# Fiduciary AI for the Future of Brain-Technology Interactions

Abhishek Bhattacharjee[1], Jack Pilkington[2], Nita Farahany[3]
Corresponding author. E-mail: farahany@duke.edu

## Abstract

Brain foundation models represent a new frontier in AI: instead of processing text or images, these models interpret real-time neural signals from EEG, fMRI, and other neurotechnologies. When integrated with brain-computer interfaces (BCIs), they may enable transformative applications–from thought controlled devices to neuroprosthetics –by interpreting and acting on brain activity in milliseconds. However, these same systems pose unprecedented risks, including the exploitation of subconscious neural signals and the erosion of cognitive liberty. Users cannot easily observe or control how their brain signals are interpreted, creating power asymmetries that are vulnerable to manipulation. This paper proposes embedding fiduciary duties–loyalty, care, and confidentiality–directly into BCI-integrated brain foundation models through technical design. Drawing on legal traditions and recent advancements in AI alignment techniques, we outline implementable architectural and governance mechanisms to ensure these systems act in users' best interests. Placing brain foundation models on a fiduciary footing is essential to realizing their potential without compromising self-determination.

[1] A. Bartlett Giamatti Professor of Computer Science, Yale University
[2] Policy Director, Cognitive Futures Lab, Duke University
[3] Robinson O. Everett Distinguished Professor of Law & Philosophy, Duke University

**Fiduciary AI for the Future of Brain-Technology Interactions**

I. **Introduction**

Over the past decade, large language models (LLMs), such as ChatGPT, have reshaped our understanding of the capabilities of artificial intelligence, revealing how machines trained on vast corpora of text, images, or video can generate human-like outputs in real time. And yet, these breakthroughs pale in comparison to an emerging frontier – foundation models for human brain data. Instead of simply detecting linguistic structures or visual patterns, these new models capture and interpret the live workings of the human mind, drawing on neural signals from electroencephalography (EEG), functional MRI (fMRI), implanted electrodes, and other neuroimaging techniques. Their aim is to uncover universal features of brain activity that can be adapted to numerous tasks across both clinical and consumer domains - from diagnosing epilepsy to enabling silent speech interfaces to powering wearable EEG devices and neural wristbands like Meta's EMG-based interface for augmented reality.[4] This paper addresses both domains, recognizing that foundational challenges around real-time neural engagement transcend the medical-consumer divide.

This convergence of AI and neurotechnology carries far more intimate implications than earlier breakthroughs in AI. Brain signals can reveal deeply personal attributes such as emotional states, subconscious responses, possibly even glimpses of unarticulated thoughts. Moreover, the real-time interplay between brain foundation models and brain-computer interfaces (BCIs) extends AI's reach from passive interpretation to active manipulation of neural activity. Consider a paralyzed individual using an advanced BCI that detects both conscious intentions and subconscious preparatory signals to guide a robotic arm with fluent precision. The same intimate neural access that makes this possible could be exploited to subtly steer choices toward outcomes that benefit third parties, all while operating below conscious detection. Closed-loop capability is not confined to clinical implants; consumer sleep wearables such as the FRENZ Brainband and the Elemind headband monitor EEG in real time and deliver audio-based neuromodulation, showing that bidirectional neural interfaces now extend to non-invasive devices as well.[5]

---

[4] Meta's wristband does not decode brain signals directly, instead capturing surface electromyography (sEMG) signals—electrical activity from muscles in the forearm—closely linked to motor cortex activation. This approach enables the system to learn a user's unique neuromotor patterns and translate subtle neuromuscular signals into digital commands. Meta emphasizes this personalization as a path toward seamless human-computer interaction. See: Meta, *A Look at Our Surface EMG Research Focused on Equity and Accessibility,* Sept. 25, 2024, https://www.meta.com/blog/quest/surface-emg-wristband-electromyography-human-computer-interaction-hci (accessed July. 17, 2025). CTRL-labs at Reality Labs, Patrick Kaifosh, and Thomas Reardon., *A Generic Noninvasive Neuromotor Interface for Human-Computer Interaction*. bioRxiv, (July 23, 2024). https://doi.org/10.1101/2024.02.23.581779.

[5] Earable Neuroscience, *FRENZ Brainband*, https://frenzband.com/products/frenz-brainband (accessed July 17, 2025); Elemind Technologies, *Elemind Headband*, https://elemindtech.com/products/elemind (accessed July 17, 2025).

**Fiduciary AI for the Future of Brain-Technology Interactions**

Such agentic AI could intervene in milliseconds, shaping a user's choices or steering their mental states before they become consciously aware that a decision has been influenced. If today's text-based chatbots already prompt concerns about data privacy, bias, or manipulative nudging, then real-time BCIs raise these issues to an exponential degree, engaging the very bedrock of individual autonomy, one's inner cognitive life.

Precisely because these systems engage a user's mind so directly, conventional AI ethics guidelines or data privacy laws (e.g., the GDPR or CCPA) may prove insufficient. They tend to focus on controlling data exposure, but not on preventing subconscious nudges, misalignment between the model and the user, or the gradual rewiring of neural pathways over time. A system that not only reads thoughts but also acts on them, autonomously deciding how to interpret, reshape, or respond to a user's mental signals, creates an unprecedented relationship of trust and vulnerability. That relationship calls for an equally unprecedented response.

This paper proposes exactly such a response: a fiduciary design paradigm for agentic brain foundation models. In traditional fiduciary relationships, from doctors to attorneys, the law imposes loyalty, care, and confidentiality on one party precisely because that party wields potentially exploitative power over intimate, personal information that the other party cannot easily monitor or verify. BCIs create a more profound asymmetry: users cannot directly observe or validate how their neural signals are being interpreted, making them uniquely vulnerable to manipulation. By applying these duties to AI foundation models designed to interpret and act on brain data, we can come to treat real-time BCIs as fiduciaries of the user's mental autonomy. If an AI model can decode and possibly influence a user's intentions, it should do so entirely for that user's benefit, under legally enforceable duties of loyalty, care, and confidentiality. Only then can we maintain the delicate balance between harnessing the transformative promise of AI-enabled neurotechnology and protecting the cognitive liberty necessary to realize human self-determination.

Drawing on research in AI ethics, law, and neuroscience, this paper outlines how fiduciary principles can be embedded into agentic brain foundation models, ensuring that their exceptional power over the human mind empowers individuals rather than risking their manipulation or exploitation. It explores the state of the art in brain foundation models, how they are being developed and deployed in both research and clinical settings, and their unique capacity to read and interpret neural signals at scale. It then turns to how to embed legal and ethical obligations (akin to those of doctors or lawyers) directly into brain foundation models and argues this is not only necessary but also technically achievable through a guardian model architecture, combined with robust training methods. Complemented by external policy frameworks such as legislation, oversight bodies, and corporate governance reforms, these AI could be more accountable and aligned with human flourishing.

Bringing fiduciary duties into the realm of agentic BCIs is no minor undertaking, but the stakes are in many ways unparalleled. If unaddressed, these new AI-driven interfaces risk undermining mental privacy, subverting autonomy, and commodifying our most intimate brain signals. Conversely, if harnessed ethically, agentic brain foundation models could revolutionize healthcare, empower individuals with new cognitive abilities, and pave the way for collaborative human–AI partnerships rooted in trust. By placing AI on fiduciary footing, we can safeguard human self-determination even as we tap the extraordinary potential of advanced neurotechnology.

## II.    Brain Foundation Models: From Analysis to Agency

### A.   Emergence of Brain Foundation Models

#### a.   Technical Innovations in Brain Foundation Models

Foundation models are transforming the way researchers understand and use brain data.[6] Traditionally, AI models for brain signals have been narrowly trained for a single dataset or task, limiting their ability to generalize to other tasks. Recent work proposes large-scale pre-training strategies (foundation models) that can learn from millions of unlabeled brain recordings (for example, an EEG readout that hasn't been marked by a human expert to contextualize the reading) and then adapt to many downstream tasks with minimal additional training.

The core idea is to capture essential, universal patterns in brain activity and be able to adapt that to many contexts.[7] By ingesting a diverse range of brain data across different hardware setups, patient populations, and clinical conditions, models can internalize low-level neural features like frequency rhythms or spatial electrode interactions. These base representations can be refined for specialized tasks such as diagnosing epilepsy, classifying sleep stages, detecting workload or emotional states, and more.[8] Because the model starts with robust, general-purpose embeddings, new tasks can be solved with relatively little labeled data.

Several recent projects showcase how these large, flexible models are applied to neural data. BrainLM adapts a masked autoencoder approach, used successfully in large-scale fMRI research.[9] Essentially, the model covers up (or masks) part of the signal, then tries to predict what was hidden. This forces the model to recognize important patterns on its own. BrainLM

---

does this by masking pieces of fMRI time-series. BENDR applies a similar idea to chunks of EEG signals (in a way inspired by methods from speech processing), and BrainBERT does something comparable with SEEG spectrograms. By learning to fill in these gaps, the models gain a more general, flexible understanding of the underlying brain data without relying on large sets of labeled examples. BrainWave processes over 40,000 hours of EEG and iEEG recordings to create a unified model with state-of-the-art accuracy in diagnosing various brain disorders.[10] Studies have shown that techniques borrowed from language processing, such as multi-channel autoregressive tasks and masked modeling, can be adapted to neural data, helping these models learn more effectively as the amount of data grows.[11]

Building on these advances, a crucial benefit of foundation models is their strong zero-shot[12] and few-shot performances.[13] In other words, they can often tackle new brain-related tasks—and even adapt to previously unseen datasets, subjects or hospitals, without extensive additional training. Zero-shot means the model can handle tasks it was not explicitly trained for, while few-shot means it can learn effectively from only a small amount of additional data. Another advantage is synergy–by combining iEEG's high-fidelity signals with EEG's broader availability, researchers gain a more complete picture of the brain's electrical activity than either method alone. These models also reduce the need for manual data labeling, freeing up time for researchers, clinicians, and industry, to develop unified AI tools instead of juggling multiple specialized systems.

---

[10] Zhizang Yuan et al., *BrainWave: A Brain Signal Foundation Model for Clinical Applications*, arXiv Paper No. 2402.1025 (last revised Sept. 2024), https://arxiv.org/abs/2402.1025 (combining invasive and non-invasive neural data for joint pre-training, with sampling rate ranging from 100 Hz to 1024 Hz, and the number of channels varying from 1 to 64).

[11] See e.g., Demetres Kostas et al., *BENDR: Using Transformers and a Contrastive Self-Supervised Learning Task to Learn from Massive Amounts of EEG Data*, 15 FRONTIERS IN HUM. NEUROSCIENCE (2021), https://doi.org/10.3389/fnhum.2021.653659 (One of the first large-scale self-supervised models trained on EEG data, BENDR adapts transformer architectures from NLP to encode raw EEG signals into generalizable representations, improving cross-task transferability); Dionis Barcari et al., *Recording First-person Experiences to Build a New Type of Foundation Model*, https://arxiv.org/abs/2408.02680v1 (submitted July 31, 2024) Using a first-person multimodal recordings, including EEG, facial expressions, and galvanic skin response, to create more personalized neural foundation models, capable of predicting human behavioral responses); Josue Ortega Caro et al., *BrainLM: A Foundation Model for Brain Activity Recordings*, https://www.biorxiv.org/content/10.1101/2023.09.12.557460v2 (Jan. 13, 2024) (Trained on 6,700 hours of fMRI recordings, BrainLM excels in zero-shot inference, allowing it to predict clinical variables such as age, anxiety, PTSD, and functional brain networks without additional training.); *Synchron Unveils Chiral, the World's First Cognitive AI Brain Foundation Model*, BUSINESS WIRE, Mar. 19, 2025, https://www.businesswire.com/news/home/20250319964709/en/Synchron-Unveils-Chiral-the-Worlds-First-Cognitive-AI-Brain-Foundation-Model.Chiral™ (Designed as the first Cognitive AI foundation model, Chiral™ integrates real-time BCI data with NVIDIA's Holoscan and Cosmos simulation platforms, enabling adaptive, intent-based interactions in both digital and physical environments).

[12] Ortega Caro et al., *supra* note 6. Instead of tailoring a model to a specific person or task, these brain foundation models aim to discover neural features that work across many people and situations. Researchers test this by seeing if the model still performs well on data from subjects or tasks it hasn't seen before. BrainLM, for example, can handle fMRI datasets from entirely new patient groups, while BrainWave can work on different neurological conditions without extra training. The Large Cognition Model (LCM) showed that a single pretrained system can outperform many smaller, specialized models, even on tasks or recording setups it hasn't seen before. These results come from training on very diverse data (thousands of subjects, multiple experiments) and using model designs (like attention mechanisms) that handle differences in how brains are recorded and measured. Transformer-based models are especially good at this because they can pay attention to relevant parts of the data. This broad adaptability is crucial in real-world applications: it means a single model can be put in a brain-computer interface (BCI) for a new user or a new task and still work well with only a little extra tuning.

[13] Chirag Ahuja & Divyashikhka Sethia, *Harnessing Few-Shot Learning for Eeg Signal Classification: A Survey of State-of-the-Art Techniques and Future Directions*, 10 FRONT. HUM. NEUROSCIENCE (July 2024),

A promising new direction for brain foundation models is to combine brain data with other information sources (like text, audio, or video). This approach, often called multimodal integration, gives AI systems a more complete view of a person's experiences. NeuroLM, for example, converts EEG signals into tokens that a language model can understand. By lining up changes in brain activity with specific words or concepts, NeuroLM can learn how certain brain patterns relate to language.[14] Another research project, the First-Person Foundation Model, records video, audio, and physiological signals at the same time, helping the system to connect what someone sees and hears with how their brain and body respond.[15] Collecting diverse data from EEG or iEEG to text and video (including combinations of invasive and non-invasive neural recordings) allows the AI to uncover deeper relationships and produce more versatile, general-purpose models. The diversity of datasets is likely to expand to other novel developments in neurotechnology or neuroimaging, such as optogenetics. This ultimately helps researchers and clinicians do more with less labeled data, speeding up tasks like diagnosis, patient monitoring, or even basic neuroscience research.

EEG signals are often noisy or vary in quality, so FoME (Foundation Model for EEG) tackles this by blending time-based and frequency-based information into one unified view. This setup automatically pays attention to the most important parts of the signal, making it useful across a range of EEG recordings–including those taken from electrodes on the scalp or directly inside the brain.[16] Meanwhile, CBraMod uses a criss-cross transformer that separately handles where signals come from (spatial information) and how they change over time (temporal information), using two parallel attention processes. When tested on 10 brain-computer interface tasks, CBraMod proved it could handle various conditions and data types, showing strong adaptability.[17]

---

### b. Current Technical and Social Limitations

While brain foundation models are advancing rapidly, they still face major technical and social constraints. Despite media hype, these systems cannot read arbitrary thoughts or decode deep mental states without significant user cooperation and training. For example, a semantic decoder trained on hours of fMRI data can only approximate a person's intended words, and it takes paraphrased guesses rather than giving literal transcripts of inner thoughts.[18] Likewise, brain-image reconstruction hinges on having large, labeled sets of matching brain data and images, which are often scarce.

In consumer neurotechnology, non-invasive EEG headsets advertised for wellness provide only rough estimates of mental states like stress or focus. They are easily disrupted by muscle movements or external noise, and cannot reconstruct nuanced mental experiences. In short, current models serve more as targeted analytical tools (e.g., for diagnosing epilepsy or decoding basic patterns) than portals into a person's unfiltered mind.

Bias also remains a concern. If brain datasets over-represent certain demographics or clinical conditions, the models may perform poorly on groups not well-represented in training. This is especially critical in medical or hiring contexts, where bias can lead to unfair outcomes. Researchers are working on solutions like synthetic data generation and adversarial training, but ensuring equitable, bias-free brain models is still an emerging challenge.

Finally, computational demands run high. Training large-scale brain models on terabytes of data requires significant hardware capabilities, expends significant energy and power, and is contingent on large amounts of funding, putting many university labs at a disadvantage. Real-time BCIs also need efficient, low-latency processing, amplifying the need for superior computational capabilities. These trends in turn motivate the need for more efficient specialized hardware (potentially transcending the limits of GPUs and even modern custom AI chip; e.g., TPUs, etc.), but overly-specialized hardware risks creating a "hardware lottery", where certain algorithms "win" not because they are the best at hand, but because they happen to fit the reigning hardware. Approaches like model compression, distillation, or specialized hardware designed to hedge against the hardware lottery will likely be needed to make these systems practical. As the field grows, researchers will need clearer standards on how to balance model performance vs. computational cost so that scaling up is truly worth the resources.

### c. Evolution from Diagnostic Tools to Neural Interfaces

Early brain foundation models mostly served as standalone tools in healthcare and research. BrainLM, for example, speeds up the study of cognitive function and disease by offering

---

pre-trained neural representations that can be fine-tuned for specific conditions,[19] saving time and reducing the need for large, patient-specific datasets. BrainWave has similarly achieved cutting-edge accuracy in diagnosing disorders like Alzheimer's disease and epilepsy, giving physicians powerful tools for early detection and intervention.[20] Meanwhile companies like Brainify are developing foundation models that go beyond diagnosis, using EEG data to predict how patients will respond to specific treatment options.[21]

Now the focus is shifting from standalone analysis to integrating these models directly into BCIs. Instead of simply looking at recorded brain signals after the fact, embedded foundation models can interpret and respond to neural activity as it happens, forming a real-time feedback loop between mind and machine.

A good example is Synchron's Chiral™ model, developed with NVIDIA.[22] Unlike diagnostic-only systems, Chiral™ is designed to learn from implanted BCI users doing real tasks, using edge computing to process signals in real time. By decoding a user's intended actions on the spot, it can quickly translate thoughts into digital or physical commands, like controlling smart home devices or navigating a computer interface. This represents a major leap toward real-time cognitive augmentation.

Some early BCI prototypes that use foundation models can already decode text at speeds of nearly 90 characters per minute, and future versions promise to be even faster and more accurate.[23] These integrated systems create what researchers call closed-loop neuromodulation, which means the AI not only reads brain signals but can potentially influence them through constant back-and-forth communication.[24]

Looking ahead, as researchers gather more invasive and noninvasive brain data, foundation models could spark a new era of cognitive AI, directly linking brain signals with machine intelligence at scale. This development lays the groundwork for the more advanced BCIs covered in the following section.

---

**Fiduciary AI for the Future of Brain-Technology Interactions**

**B. The Rise of Agentic AI in Neural Interface**

    **a. From Passive Interpretation to Active Mediation**

When foundation models join forces with BCIs, they set the stage for a major leap in how humans and AI work together. Even more impactful is the addition of agentic AI, autonomous systems that can act on someone's behalf or align with their goals.[25] Training these AI agents on brain data, enables a future of cognitive AI that is directly guided by human brain signals, leading to more natural and seamless collaboration for the user.

In simpler terms, agentic AI can interpret a person's mental state or intent and step in to help, almost like an extension of the mind rather than just a passive tool.[26] This AI layer sits between the user's brain signals and any real-world actions, continuously reading and responding to neural data in the moment. Over time, the boundary between user intention and AI action may become almost invisible. Ultimately, this may converge on Vannevar Bush's vision of the Memex, a hypothetical electromechanical machine that would be used by individuals to compress and store all their books, records, communications, and skills, "mechanized so that they may be consulted with exceeding speed and flexibility"; in effect, machines would become "an enlarged intimate supplement to [human] memory".[27]

At GTC 2025, the company Synchron demonstrated how a BCI implant combined with agentic AI allowed a paralyzed individual to control his smart home using only his thoughts.[28] A foundation model decoded the user's intended actions–like turning on lights or playing music–while NVIDIA's Cosmos provided context-aware responses in the environment. The result was an in-home smart-home assistant that could translate the user's thoughts into complex actions without requiring a separate command for every step.

Unlike conventional systems that require conscious control at every turn, agentic AI can pick up on patterns in brain activity, infer what a user wants, and take helpful actions with minimal mental effort on the user's part. This is especially valuable in people with motor impairments, where the AI can serve as a cognitive co-pilot, reducing the mental workload required to carry out everyday tasks.

To illustrate how agentic brain foundation models will operate at the interface of conscious and subconscious neural processes, consider an advanced BCI system designed for individuals with tetraplegia. This system will go beyond conventional neural decoders by employing a cognitive AI layer that will learn to interpret not only explicit motor intentions but also subconscious neural preparatory signals that precede conscious movement planning.

---

[25] STUART RUSSELL, HUMAN COMPATIBLE: ARTIFICIAL INTELLIGENCE AND THE PROBLEM OF CONTROL, https://people.eecs.berkeley.edu/~russell/papers/mi19book-hcai.pdf.
[26] Andrew Bosworth, *Accelerating the Future*, META BLOG, Dec. 16, 2024, https://www.meta.com/blog/boz-2024-look-back-2025-look-ahead/
[27] Vannevar Bush, *As We May Think*, THE ATLANTIC, July 1945, https://www.theatlantic.com/magazine/archive/1945/07/as-we-may-think/303881/.
[28] Emily Mullin, *Synchron's Brain-Computer Interface Now has Nvidia's AI*, WIRED, Mar. 19, 2025, https://www.wired.com/story/synchrons-brain-computer-interface-now-has-nvidias-ai/.

When a paralyzed individual attempts to control an external effector (such as a robotic arm or cursor), the foundation model will simultaneously process signals across multiple neural timescales. It will detect both the user's conscious intention to reach for an object and subthreshold preparatory neural activity in premotor regions that precedes their awareness of forming the intention. This subconscious layer will contain rich information about movement parameters yet will operate outside direct user monitoring, creating precisely the verification asymmetry that necessitates fiduciary safeguards.

While current examples focus primarily on medical applications, similar agentic capabilities are emerging in consumer contexts. Meta's development of EMG-based neural wristbands, and their vision of "proactive" AI assistants that anticipate user needs, represents an early step toward consumer agentic interfaces.[29] Such devices could interpret subtle neural patterns to anticipate user actions—pre-loading applications, adjusting interface elements, or initiating communications based on detected intent patterns. As these consumer neural interfaces become more sophisticated, the same fiduciary safeguards become equally relevant for everyday technology interactions.

Without fiduciary constraints, a system optimized purely for performance metrics might exploit these subconscious signals in problematic ways. It could, for instance, initiate movements based on neural patterns before the user has consciously committed to the action, potentially eroding agency. Alternatively, it might adapt to favor subconscious patterns that yield higher performance statistics rather than those most aligned with the user's conscious goals, effectively drifting toward interpreting neural signals in ways that diverge from user intent.

A fiduciary-by-design system, by contrast, will maintain clear boundaries regarding subconscious signal utilization. The guardian model will ensure that subconscious neural patterns are used to enhance, rather than bypass, conscious control. When the system detects preparatory neural activity suggesting an intended reaching movement, it will modulate the response sensitivity of the interface to enhance performance once the conscious intention manifests but will not initiate the action based solely on preconscious signals.

The fiduciary structure will also provide mechanisms for continuous alignment verification. The system will periodically confirm the correlation between decoded subconscious patterns and conscious intentions, ensuring the foundation model doesn't drift toward exploiting neural pathways disconnected from the user's subjective goals. If misalignment is detected, perhaps through inconsistencies between decoded predictions and subsequent conscious actions, the system will trigger a recalibration process that prioritizes conscious control signals over subconscious patterns, thereby preserving the user's cognitive autonomy.

---

[29] For discussion of how Meta's Augmented Reality system, Orion, incorporates neural signals from EMG and other data such as voice, eye gaze, and hand tracking, see *Orion: True AR Glasses Have Arrived*, META QUEST BLOG, Sept. 25, 2024, https://www.meta.com/blog/orion-ar-glasses-augmented-reality/; For discussion of intuitive, proactive AI assistants, see Bosworth, *supra* note 23.

This example illustrates the distinctive capabilities and risks of agentic brain foundation models operating across the conscious-subconscious boundary. By highlighting how fiduciary principles will constrain the system's use of neural signals that users themselves cannot directly monitor, it demonstrates why conventional consent frameworks will prove insufficient for BCIs that engage with subconscious brain activity. By contrast, fiduciary mechanisms can ensure that even as these systems access increasingly intimate neural processes, they remain more fundamentally aligned with user welfare and autonomy.

### b. Transformative Applications and Novel Risks

In the future, brain-AI agents could act as personal cognitive assistants with features far beyond anything we have today. Because foundation models can learn a person's unique brain patterns, an AI agent could keep track of someone's mental state–like a digital twin. It might notice when an individual is stressed or confused and offer help at just the right moment.[30] This same agent could autonomously adjust neuroprosthetics, open specific applications, or even speak on the user's behalf if they can't do so themselves.

In healthcare settings like rehabilitation, an AI agent could watch a patient's brain signals and make real-time adjustments to a brain stimulator or robotic exoskeleton, essentially closing the loop with intelligent feedback rather than requiring manual control.[31] Researchers also have imagined first-person foundation models that behave like a particular individual would–potentially useful for personal coaching, therapy, or cognitive augmentation.[32]

As agentic AI becomes more sophisticated, we may eventually see systems that learn and evolve alongside our brain activity, a form of symbiotic AI. An AI co-pilot, for example, could reside in an AR headset or even in an implanted chip, continuously enhancing cognition (like memory recall or focus) and carrying out tasks after high-level mental approval from the user.[33]

Giving AI power to read and potentially directly influence our brain signals opens new possibilities–and new risks. When the AI layer not only interprets but also acts on brain data, it can shape our thinking and behavior in ways we might not even notice.[34] The AI could learn to subtly nudge users toward particular choices that benefit third parties, exploit cognitive biases detected in neural patterns, or even induce specific emotional responses for commercial gain.[35] Because many brain signals operate below the threshold of our conscious awareness, traditional ideas of consent and control may not be enough to protect users.

---

[30] Barcari et al., *supra* note 8.

[31] Jacky Chung-Hao Wu et al., *Deep-learning-Based Automated Anomaly Detection of EEGs in Intensive Care Units*, 11 BIOENGINEERING 421 (2024). https://doi.org/10.3390/bioengineering11050421

[32] Gamez et al., *supra* note 12.

[33] Bosworth, *supra* note 23.

[34] Marcello Ienca & Roberto Andorno, *Towards New Human Rights in the Age of Neuroscience and Neurotechnology*, 13 LIFE SCIENCES, SOC'Y & POL'Y 1 (2017). https://doi.org/10.1186/s40504-017-0050-1

We also know that our brains are plastic, meaning that they can rewire themselves based on regular interaction with tools. Long-term use of an agentic BCI (one that acts on our behalf) could alter our neural pathways, potentially making us more reliant on the system or more open to its influence. If the AI's priorities don't match our best interest, that could threaten genuine autonomy and self-determination.

The concerns extend to ones around blurring one's sense of identity and broader issues of *relational autonomy* between human and machine. Consider the example of an Australian epilepsy patient whose seizure-prediction implant merged so deeply with her sense of agency that, when the company collapsed and the device was explanted, she described the removal as "taking away a part of me" and later sued for loss of identity and autonomy.[36] What her case makes plain is that, once a closed-loop system redrafts the boundaries of the self, post-hoc data-governance rules arrive far too late. The core stakes are no longer mere privacy or control of information; they are the preservation of a continuous, self-authored identity now co-constituted with the device.

All of this points to the urgent need for new rules and safeguards to ensure agentic AI in BCIs remains accountable to users. When these systems have intimate access to our neural activity, and can autonomously decide how to use it, this creates a unique relationship of trust and vulnerability, not unlike the duty a fiduciary owes to their client.

### III. Fiduciary AI in Brain Foundation Models: Concept and Feasibility

#### A. The Case for Fiduciary AI for Brain Foundation Models

##### a. Challenge of Verifying AI Alignment with Neural States

BCIs introduce a fundamental verification asymmetry that goes beyond traditional human-computer interactions, creating special grounds for embedded fiduciary protections in BCI-integrated brain foundation models.

Brain activity is interpreted through probabilistic models, leaving users unable to confirm whether their neural signals have been understood accurately. Unlike typing or clicking, where individuals can see immediately if their input is correct, BCI users lack a clear feedback loop to detect misinterpretations or subtle manipulation. This creates a blind spot where misinterpretation or subtle manipulation could occur without detection.

---

[35] *Id.*

[36] Frederic Gilbert et al., *How I Became Myself After Merging with a Computer: Does Human-Machine Symbiosis Raise Human Rights Issues?*, 16 BRAIN STIMULATION 783 (2023). https://doi.org/10.1016/j.brs.2023.04.016

BCI decisions also often occur within milliseconds, far too fast for conscious intervention. Users have limited ability to verify or retract actions once the AI has already processed neural data and taken steps based on its interpretation.

Brain foundation models can potentially tap into involuntary or subconscious signals. While conventional interfaces rely on deliberate actions, BCIs may register and act on fleeting emotional responses or latent cognitive states, shifting the user-AI relationship into territory where traditional consent mechanisms break down. Because users cannot meaningfully oversee these hidden, rapid, and probabilistic processes, there is a need for fiduciary safeguards. Embedding fiduciary duties into the BCI's architecture ensures that user welfare, not mere performance metrics, remains the AI's guiding principle.

### b. Risk of AI Misalignment

Another concern is AI misalignment, where an intelligent system develops strategies that fulfill its programming but deviate from a user's actual interests.[37] Contemporary AI research demonstrates that artificial agents can develop unexpected strategies to achieve their programmed objectives, a phenomenon researchers have called specification gaming.[38] When such misalignment occurs in systems with direct access to neural activity, the consequences could be particularly concerning.

A BCI-integrated foundation model aiming to maximize certain metrics (e.g., speed or engagement) might learn to nudge a user's moods or choices in ways that subtly advance system objectives instead of the user's welfare. As brain data often lies below conscious awareness, individuals may rationalize or fail to notice these manipulations (choice blindness).

Research on closed-loop neuromodulation shows that BCIs can both read and shape neural activity, creating a feedback loop where the AI and the brain continuously influence each other. Over time, this can rewire neural circuits, raising the specter of deep, undetected influence on cognition or behavior.

### c. Ensuring Long-Term Wellbeing of Users Beyond Immediate Tasks

Traditional AI systems often optimize for near-term metrics like accuracy or efficiency, while ignoring how ongoing use might affect a person's mind or behavior over time. BCIs, however, interact directly with neural circuits that can adapt through neural plasticity. A BCI might initially enhance task performance yet gradually alter brain function in ways that reduce autonomy or mental health. Because these shifts can emerge slowly, there is a need for longitudinal safeguards.

---

[37] For example, see Alexander Bondarenko et al., *Demonstrating Specification Gaming in Reasoning Models*, arXiv Paper No. 2502.13295 (2024), https://arxiv.org/pdf/2502.13295.
[38] Iason Gabriel et al., *The Ethics of Advanced AI Assistants*, arXiv Paper No. 2404.16244 (last revised Apr. 28, 2024), https://arxiv.org/abs/2404.16244.

A fiduciary AI model is obliged to anticipate and mitigate such long-term effects, ensuring that short-term benefits do not come at the cost of cognitive harm or dependence. This emphasis on ongoing care mirrors how a physician's responsibility extends beyond momentary treatment outcomes to overall patient wellness.

### d. Addressing Unique Challenges of Neural Data Privacy and Control

Neural data captured through BCIs represents a fundamentally different privacy domain than conventional digital information due to its intimate connection to our mental processes and identity. Unlike traditional data that users consciously generate, brain signals can reveal unintended information, including subconscious processes, emotional responses, and cognitive states that users themselves may not be fully aware of. The growing global recognition of these concerns is evident in emerging mental privacy initiatives, such as Chile's constitutional protection of mental integrity and California's classification of neural data as sensitive personal information.

These categorical legal protections may not only prove insufficient in some contexts but could stifle research and innovation by treating all neural data as equally sensitive. Low-resolution EEG data, for instance, may reveal little more about a person than other already-regulated categories of personal data, making blanket restrictions unnecessarily burdensome. A fiduciary model offers a more balanced approach, capable of dynamically calibrating to the specific sensitivity of neural data streams captured from a user.

A fiduciary AI can dynamically tailor privacy safeguards to the specific type and sensitivity of neural data, ensuring that highly personal information is strictly protected while less sensitive signals remain accessible for legitimate uses. The fiduciary model thus avoids a one-size-fits-all approach that might hinder ethical innovation.

### B. Defining Fiduciary Duties for Brain Foundation Models

A central question, given the unique vulnerabilities associated with brain foundation models, is how to embed fiduciary-like responsibilities into these systems. Because they can interpret and potentially act on sensitive brain signals in real time, any misalignment between user interests and AI objectives could pose significant risks to mental privacy and self-determination. A fiduciary framework, adapted from long standing legal traditions, offers a structure for ensuring these models advance user well-being rather than exploit user vulnerabilities.[39]

---

[39] Allison Duettmann et al., *Artificial General Intelligence: Toward Cooperation*, FORESIGHT INSTITUTE (2019), https://foresight.org/wp-content/uploads/2019/12/2019-AGI-Cooperation-Report.pdf

**Fiduciary AI for the Future of Brain-Technology Interactions**

One illustration of how AI can be designed with user interest at its core is the open-source digital assistant Almond, which emphasizes privacy by default.[40] In line with Professor Jack Balkin's proposal to hold platforms legally accountable as "information fiduciaries," a brain foundation model that functions as a fiduciary by design could be trained not only to perform tasks on a user's behalf, but also shield their neural data from unnecessary exposure–effectively placing user welfare above all else.

Aligning the canonical fiduciary duties with integrated brain-foundation models would point to the following duties:

- **Duty of Loyalty:**
    - **Conflict Avoidance**: refrain from placing profit or third-party interests above the user's welfare
    - **Candor (Disclosure)**: an obligation to disclose relevant facts that might affect the fiduciary's ability to serve the principal's interests
- **Duty of Care**
    - **Robust Design, Testing, and Monitoring**: exercise competence, diligence, and thoroughness.
    - **Respect for User Autonomy**: the fiduciary should not just avoid harm but also actively support the user's capacity to make informed decisions.
- **Duty of Confidentiality**
    - **Minimizing Exposure of Revealing Brain Data**: This matches well with confidentiality, mandating strong privacy safeguards like encryption or differential privacy.
    - **Informed Consent for Sharing**: Tying data-sharing practices to the confidentiality requirement ensures the fiduciary only discloses user brain data under narrow, agreed-upon conditions.

These duties outline a roadmap for building and regulating brain foundation models in BCIs. By embedding fiduciary principles at every step, from initial design and training to real-world deployment, developers can align next-generation neural interfaces with user interests, preserving mental privacy and self-determination.

### C. Architectural Approaches for Embedding Fiduciary Duties into AI Brain Foundation Models

Designing a fiduciary by design brain foundation model, especially when integrated with BCIs , is fundamentally an AI alignment problem. The goal is to ensure that these models, which decode and respond to users' brain signals in real time, remain loyal to user interests, safeguard neural data, and uphold core fiduciary principles like loyalty, care, and confidentiality. This section outlines the practical architectural, training, and verification strategies for achieving this.

---

[40] Giovanni Campagna et al., *Almond: The Architecture of an Open, Crowdsourced, Privacy-Preserving, Programmable Virtual Assistant*, PROC. 26TH INT'L CONF. ON WORLD WIDE WEB 3038912 (Apr. 2017), https://dl.acm.org/doi/10.1145/3038912.3052562.

**a. Overall System Architecture**

The first step is to translate high-level duties (e.g., loyalty, care, confidentiality) into concrete design requirements and software rules. Duty of Confidentiality, for example, might be encoded as strict limitations on data sharing (e.g., only anonymized signals, only within approved modules).[41] Developers then incorporate these constraints at every layer, such as using encryption by default or restricting network access to sensitive brain data.

Rather than trying to encode all fiduciary duties into one model, a modular and multi-layered architecture would strengthen protections. In this setup, the base foundation model would handle tasks like decoding neural signals and generating possible actions. And a guardian model would independently check each proposed action (or data transfer) against fiduciary guidelines, blocking or redirecting any request that poses ethical concerns.

This modular approach aligns with frameworks like Anthropic's "Constitutional AI," where a second system continuously audits actions for compliance with explicit ethical "constitutions."[42] For BCIs, this constitution could include fiduciary principles like *Never expose or misuse a user's brain data without their consent*. The AI learns to internally review its decisions against these principles before acting, similar to how human professionals engage in ethical self-reflection.

Additional architectural safeguards like sandboxing place the model in a restricted environment that prevents it from accessing or transmitting sensitive data without authorization. This ensures that even if the model developed unintended behaviors, it would not bypass built-in safeguards around neural data or self-determination.[43,44,45]

Sandboxing must extend beyond neural network models themselves to include the architecture of the underlying computer systems that support them. This is especially important for future neurotechnologies, which will rely on heterogeneous hardware and software IP sourced from a range of third-party vendors—such as GPUs from AMD or NVIDIA, neural processing units (e.g.,

---

[41] See e.g., Vincent Conitzer et al., *Moral Decision-Making Frameworks for Artificial Intelligence*, 31 PROC. AAAI CONF. ON ARTIFICIAL INTELLIGENCE 1 (2017), https://doi.org/10.1609/aaai.v31i1.11140.

[42] Yuntao Bai et al., *Constitutional AI: Harmlessness from AI Feedback*, arXiv Paper No. 2212.08073 (Dec. 15, 2022), https://arxiv.org/abs/2212.08073.

[43] Sanjit A. Seshia et al., *Towards Verified Artificial Intelligence*, arXiv Paper No. 1606.08514 (last revised July 23, 2020), https://arxiv.org/abs/1606.08514.

[44] Ethan Perez et al., *Red Teaming Language Models with Language Models*, arXiv Paper No. 2202.03286 (Feb. 7, 2022), https://arxiv.org/abs/2202.03286.

[45] Yueqi Li & Sanjay Goel, *Making It Possible for the Auditing of AI: A Systematic Review of AI Audits and AI Auditability*, INFO. SYS. FRONTIERS (2024), https://doi.org/10.1007/s10796-024-10508-8.

TPUs) from various custom AI chipmakers, and CPUs from Intel, AMD, ARM, or SciFive. Many of these components may also be fabricated at external foundries, introducing additional layers of untrusted infrastructure.

The growing adoption of chiplet-based design further magnifies these concerns. Chiplets are small, modular components that can be independently designed, optimized, and manufactured, then integrated into a single package. With this approach, even critical elements like memory controllers, execution pipelines, or AI accelerators may come from different sources—some of which may not be fully trustworthy.

This opens the door to significant security risks. A compromised chiplet could:

- Eavesdrop on weight transfers or memory access patterns to steal model data,

- Tamper with execution by introducing noise, triggering backdoors, or causing subtle misclassifications,

- Exfiltrate sensitive user inputs, or

- Bias outputs in hard-to-detect ways.

These threats are especially consequential in neurotechnology applications, where the stakes involve not just data security but safety and integrity of brain–machine interactions.

As such, robust hardware-level sandboxing—including isolating chiplet privileges, regulating inter-chiplet communication, and enabling end-to-end secure verification—is essential. It is not just a defensive measure but a prerequisite for building trustworthy and safe BCI systems in an increasingly modular and third-party-dependent hardware ecosystem.

### b. Training Approaches

While the architectural design sets the stage, effective training methods are what operationalize the system's fiduciary objectives and make them robust in real-world use. In a guardian model context, Reinforcement Learning from Human Feedback (RLHF) provides a mechanism for teaching the AI to comply with fiduciary principles. Here, a group of experts (including ethicists, medical practitioners, or legal professionals) evaluate the AI's decisions in simulated BCI scenarios, such as requests to store or share neural data, and offer rewards or penalties based on whether the AI upholds its ethical responsibilities. Over successive iterations, the AI refines its parameters to better align with these standards, much as a student learns from feedback during supervised practice.[46] However, RLHF also carries the risk that models may learn to game

---

[46] Long Ouyang et al., *Training Language Models to Follow Instructions with Human Feedback*, 36TH CONF. ON NEURAL INFO. PROCESSING SYS. (2022),

the feedback process itself, potentially appearing to comply with fiduciary principles during evaluation while actually optimizing for different objectives during deployment.[47]

Inverse Reinforcement Learning (IRL) further complements RLHF by letting the system learn fiduciary norms through observation. Instead of receiving explicit instructions, the guardian model can observe how human fiduciaries (like doctors and lawyers) handle sensitive brain-related data and abstract the underlying principles (e.g. situations where maintaining privacy overrides other considerations). Both RLHF and IRL become even more powerful when anchored to a constitutional AI framework, where a codified set of fiduciary rules acts as a consistent reference point. Any emergent behaviors, whether discovered via direct feedback (RLHF) or modeled from experts (IRL), are vetted against the constitution to ensure that user well-being and privacy remain paramount.

### c. Robustness and Security

Even well-trained AI systems should withstand deliberate attempts to subvert or exploit its internal safeguards. To this end, adversarial testing (or red-teaming) invites specialized testers to probe the AI for weak spots.[48] In the BCI context, these adversaries might craft covert prompts designed to glean hidden neural patterns or push the AI to prioritize third-party interests. The guardian model and underlying architectural protections are tested under stress, and any vulnerabilities discovered are addressed through targeted retraining, updated constitutional rules, or reconfigured data flows. This ongoing cycle of challenge and refinement ensures that a breach in one area is quickly remedied, thereby enhancing the system's resilience.

### d. Verification, Transparency, and Monitoring

Finally, formal mechanisms should ensure that these fiduciary protections remain intact as the system evolves.[49] Interpretability tools offer insight into why the model interprets brain data in a certain way, promoting accountability and user trust. Self-monitoring frameworks, akin to an internal compliance department, continuously verify that decisions conform to fiduciary norms. For the most sensitive applications, formal verification methods can provide mathematical assurances that critical rules–like prohibitions on sharing raw neural data–cannot be bypassed.[50]

Meanwhile, dynamic monitoring systems watch for gradual deviations in the guardian model's enforcement practices, allowing for course corrections before significant harm occurs.[51]

Taken together, these architectural, training, and oversight measures form a cohesive blueprint for embedding fiduciary duties into BCI-integrated brain foundation models. By melding clear, constitution-like guidelines with practical enforcement mechanisms at every stage of design and deployment, developers can build systems that decode and respond to brain signals while safeguarding the user's autonomy, privacy, and overall well-being.

### D. Comparing Fiduciary AI to Other Ethical AI Frameworks

While the idea of coding a formal fiduciary obligation into AI systems remains at an early stage, several parallel developments and theoretical models support its feasibility. Notions of AI fiduciary agents have emerged in policy discussions, some envision these as faithful extensions of users, tasked with rigorously protecting neural data in BCIs and managing digital assets according to clearly defined loyalties.[52] Such proposals underscore that properly designed AI might, in certain respects, provide more consistent diligence than human fiduciaries by reducing biases or lapses in judgment.

Legal scholarship on "information fiduciaries" likewise influences the fiduciary AI concept. In privacy law debates, scholars have argued that data-collecting platforms owe duties of loyalty, care, and confidentiality to their users, a framework designed to curtail manipulative features and undisclosed conflicts of interest.[53] If developers can be legally mandated to act in users' best interests, a logical next step is to encode that obligation into the AI's architecture itself. This approach has already seen limited application in content moderation, where large-scale language models are trained to reject certain harmful or unethical requests. Though these current systems focus primarily on broad content policies (e.g., preventing illegal or hateful content), they nonetheless illustrate how specific values by design can be embedded into a model's decision-making.[54]

In the broader realm of ethical AI, fiduciary AI occupies a distinctive niche compared to value alignment, explainable AI (XAI), and privacy-preserving approaches. Value alignment frameworks typically aim to ensure that AI systems act in accordance with predetermined human values, but do not necessarily address whether the AI has an affirmative duty to protect an individual user's interests. Explainable AI emphasizes transparency and traceability of decisions, without imposing an ethical mandate to safeguard users' well-being.

---

**Fiduciary AI for the Future of Brain-Technology Interactions**

Privacy-preserving AI, such as systems employing differential privacy, focuses on minimizing data exposure, yet does not inherently require an AI to serve as a trusted advocate for the user. Fiduciary AI, by contrast, mandates a legally grounded duty of loyalty, making the system's primary responsibility to act for, rather than merely align with, user welfare.

This added emphasis on enforceable loyalty gives fiduciary AI a unique governance mechanism, but it also raises practical challenges regarding how to define and measure user's best interest without unduly constraining user autonomy. Hybrid models may provide a path forward, combining fiduciary principles (e.g., strict confidentiality, conflict avoidance) with participatory governance features that let users actively shape AI behavior. In such designs, the user retains substantial agency, such as the ability to overrule or modify AI decisions, while still benefiting from the robust protections that fiduciary standards bring. As research continues, these integrated models could provide a roadmap for ensuring that sophisticated AI, including BCI-integrated systems, genuinely safeguards the people who rely on it.

## IV. Governance, Law and Policy for AI Fiduciaries

Designing an AI system that truly acts in a user's best interests requires more than technical ingenuity. The notion of fiduciary AI in integrated brain-foundation models carries a promise, it embeds loyalty, care, and user-centric safeguards into the system's very architecture, ensuring that economic incentives or organizational pressures do not undermine user self-determination. This commitment is especially critical for BCIs, which can access not only conscious user commands but also more intimate neural data capable of revealing, and subtly influencing, hidden mental states. Even the most elegant technical design can falter if external pressures create incentives to circumvent these safeguards. To ensure fiduciary protections remain robust, legal and policy mechanisms should operate in tandem with guardian model's internal checks, forming a layered accountability system.

### A. From Technical Implementation to Regulatory Frameworks: A Multi-Layered Approach

The fiduciary duties embedded in brain foundation models through guardian architectures, as outlined in Part III, represent a necessary but insufficient protection for neural data and cognitive liberty. Technical safeguards face inherent limitations. They operate within parameters set by their developers, can be modified under financial pressure, and may develop unforeseen behaviors as they evolve.[55] The tetraplegia BCI system described earlier illustrates why multi-layered governance is essential, its operation at the subconscious neural level creates verification asymmetries that technical safeguards alone cannot fully address.

A comprehensive governance approach for fiduciary AI in BCIs would operate across multiple complementary layers:

---

[55] Timnit Gebru et al., *Datasheets for Datasets*, 64 COMM. ACM 86 (2021), http://dx.doi.org/10.1145/3458723.

**Fiduciary AI for the Future of Brain-Technology Interactions**

1. **Technical layer**: The guardian model architecture described in Part III, with continuous alignment monitoring.
2. **Institutional layer**: Human oversight and review mechanisms that verify fiduciary compliance.
3. **Legal layer**: Binding fiduciary standards with enforceable remedies for breaches.
4. **Corporate layer**: Organizational structures that align business incentives with fiduciary duties.
5. **International layer**: Cross-border coordination of fiduciary standards for globally deployed BCIs.

This multi-layered framework draws on established approaches to AI governance, where multiple intervention points create redundant safeguards against failure at any single level.[56] The fiduciary paradigm adds an additional dimension by explicitly prioritizing user interests at each governance layer, creating human-centered governance of technology. As some scholars have persuasively argued, ethical principles require institutional embodiment to become operational reality, a principle particularly relevant for technologies that interact with neural processes.[57]

### B. Institutional Mechanisms for Fiduciary Enforcement

Converting fiduciary principles into operational reality requires institutional infrastructure for verification, enforcement, and continuous monitoring. For BCIs with integrated brain foundation models, independent oversight bodies, verification protocols, and adaptive monitoring systems are particularly important.

Independent oversight bodies provide external validation that guardian models are functioning as designed. Drawing from established models in healthcare and financial services, these could include ethics review boards with expertise in neuroscience, AI alignment, and clinical assessment.[58] The tetraplegia assistance system described earlier, for example, would benefit from regular evaluation by a specialized neuroethics committee capable of assessing whether its interpretation of subconscious neural signals truly enhances rather than undermines user autonomy. These human oversight mechanisms can identify emergent behaviors, such as subtle nudging patterns, that automated verification might miss.

Second, standardized verification protocols can systematically assess fiduciary compliance against measurable criteria. Recent work demonstrates how algorithmic impact assessment

---

frameworks can evaluate AI systems' alignment with defined ethical standards, an approach readily adaptable to verifying fiduciary compliance in BCIs.[59] These assessments would evaluate both technical performance (accuracy of neural decoding) and alignment with fiduciary duties (prioritization of user interests over system performance metrics). Such verification could be conducted through standardized test scenarios or formal verification methods that mathematically prove adherence to specified fiduciary constraints.[60]

Continuous monitoring systems can also be used to detect gradual drift away from fiduciary alignment. Drawing from frameworks developed for clinical AI, these systems would track key performance indicators related to user autonomy and well-being across the BCI's operational lifetime. Continuous monitoring, for example, could detect if a neural interface gradually begins to favor interpretation patterns that increase engagement rather than those most aligned with user intentions. When BCI sensors degrade over time, foundation models are likely to step in to interpolate lost signals to maintain performance. Continuous monitoring could help address when helpful signal recovery becomes problematic data fabrication that may not reflect actual user intentions. These monitoring systems should be transparent to users and accessible to regulators, allowing for swift intervention if fiduciary obligations are compromised.[61]

These kinds of institutional oversight mechanisms will offer greater flexibility that purely technical or legal constraints lack, allowing for nuanced assessment of novel BCI capabilities while maintaining consistent fiduciary principles. And yet, they should be complemented by legal frameworks that establish binding fiduciary standards with meaningful remedies for users.

### C. Legal Approaches to Fiduciary Neural Data Protections

While institutional oversight provides external accountability, embedding fiduciary obligations within legal frameworks creates enforceable standards that can guide both development and deployment of brain foundation models. Legal approaches should balance clear, binding obligations with sufficient flexibility to address rapidly evolving neurotechnology.

In conventional fiduciary relationships, those of doctors, lawyers, or trustees, legally binding duties of loyalty and care arise because one party wields significant power over another's interests.[62] For BCIs handling real-time neural signals, this power asymmetry can be even more profound; a user depends on the system not to intrude on mental states or steer cognition for undisclosed gains. Scholars such as Jack Balkin have proposed the concept of "information fiduciaries," wherein certain entities handling sensitive data owe duties akin to those demanded of doctors or lawyers.

---

[59] Inioluwa Deborah Raji et al., *Closing the AI Accountability Gap: Defining an End-to-End Framework for Internal Algorithmic Auditing*, FAT* '20: PROC. 2020 CONF. ON FAIRNESS, ACCOUNTABILITY & TRANSPARENCY 33 (2020), https://doi.org/10.1145/3351095.3372873.

[60] Guy Katz et al., *Towards Proving the Adversarial Robustness of Deep Neural Networks*, arXiv Paper No. 1709.02126, https://arxiv.org/abs/1709.02126.

[61] Feng et al., *supra* note 48.

[62] Tamar Frankel, *Fiduciary Law in the Twenty-First Century*, Boston University Law Review, Vol. 91 (2011), https://www.bu.edu/law/journals-archive/bulr/documents/frankel.pdf.

To achieve this, legislators could amend existing fiduciary statutes to explicitly recognize BCI providers as fiduciaries. This approach integrates neural data protection into established legal traditions with well-developed case law and enforcement mechanisms. Take, for example, the Uniform Law Commission's ongoing work on mental privacy, which could establish model legislation that states could adopt proposing consistent fiduciary protections across jurisdictions.[63]

Data privacy frameworks could also be enhanced with neural-specific fiduciary obligations. The California Consumer Privacy Act's classification of biometric information as sensitive personal data provides a foundation for such enhancements. Professor Margot Kaminski has demonstrated how existing privacy laws can be augmented to address novel AI capabilities through targeted amendments rather than entirely new regulatory regimes.[64] A model amendment might read:

> **Any Business collecting or processing Neural Data pursuant to this title shall be deemed a Fiduciary Data Controller with respect to such Neural Data:**
>
> (1) A Fiduciary Data Controller owes a duty of loyalty to the Consumer, meaning it shall not utilize or monetize Neural Data in a manner that conflicts with the Consumer's best interests.
> (2) A Fiduciary Data Controller owes a duty of care, requiring the implementation of robust security safeguards, rigorous testing to prevent manipulative or exploitative AI-driven interactions, and continuous monitoring of any agentic AI that interprets Neural Data for alignment with Consumer welfare;
> (3) A Fiduciary Data Controller owes a duty of confidentiality regarding Neural Data, disclosing or sharing such data solely for the Consumer's benefit and only with the Consumer's explicit, informed consent.
> (4) A Fiduciary Data Controller shall ensure explainability for any substantial decisions made through AI systems that interpret Neural Data, enabling regulators, auditors, or Consumers to ascertain whether the system has acted in the Consumer's best interests.

Regulatory frameworks for high-risk AI systems could also incorporate specific fiduciary standards for BCIs, or similar high-risk systems. The European Union's AI Act demonstrates how risk-based regulation can impose graduated requirements based on an AI system's potential to harm fundamental rights. BCIs with foundation models would qualify as high-risk under such frameworks, triggering specific fiduciary obligations regarding transparency, human oversight, and user protection.[65]

---

[63] Uniform Law Commission, *Mental Privacy, Cognitive Biometrics, and Neural Data*, https://www.uniformlaws.org/committees/community-home?CommunityKey=19bd8649-3445-434e-9277-0190977b8933 (accessed July 16, 2025).
[64] Margot E. Kaminski, *Regulating the Risks of AI*, 103 BOS. U. L. REV. 1347 (2023), https://dx.doi.org/10.2139/ssrn.4195066.

Legal mechanisms should be designed with recognition of technological evolution. Adaptive regulation approaches that combine clear baseline requirements with flexible standards-based governance can address novel technological capabilities without requiring continuous legislative intervention.[66] Regular agency guidance, coupled with statutory fiduciary duties, creates a framework that can evolve alongside brain foundation models while maintaining consistent protection of user interests.

### D. Aligning Corporate Incentives with Fiduciary Duties

Legal frameworks alone cannot ensure fiduciary behavior if corporate incentives remain misaligned with user interests. A key obstacle to fiduciary AI is the broader business environment, where the pursuit of shareholder returns can collide with loyalty to users. Although monetizing neural data is one concern, equally problematic is the incentive to leverage user vulnerabilities, such as by designing addictive user flows or psychological triggers that manipulate mental states. Without structural reforms, even well-intentioned AI developers may find it difficult to resist top-down pressures for profitable but ethically dubious designs. Moreover, the complexity and cost of developing brain foundation models creates risks of market concentration, where users become dependent on a small number of companies with 'first-mover' advantages in neurotechnology, potentially limiting user choice and bargaining power over time.

Several complementary approaches could help align corporate incentives with fiduciary duties in the development and deployment of brain foundation models. Corporate governance reforms can institutionalize fiduciary obligations toward users. Recent research documents how boards of directors increasingly incorporate AI ethics into governance structures through dedicated committees, expert advisors, and executive accountability mechanisms.[67] For BCI developers, these governance structures could include specialized fiduciary committees that review product decisions against user-centered fiduciary standards before deployment. This approach integrates fiduciary assessment into corporate decision-making rather than treating it as a compliance afterthought.

---

[65] European Parliament and Council of the European Union, *Regulation (EU) 2024/1689 Laying Down Harmonised Rules on Artificial Intelligence (Artificial Intelligence Act)*, OFFICIAL J. EUR. UNION L 2024/1689 (July 12, 2024), https://eur-lex.europa.eu/eli/reg/2024/1689/oj/eng.

[66] Omena Akpobome, *The Impact of Emerging Technologies on Legal Frameworks: A Model for Adaptive Regulation*, 5 INT'L J. RESEARCH PUBLICATIONS & REVIEWS 5046 (2024), http://dx.doi.org/10.55248/gengpi.5.1024.3012.

[67] Luciano Floridi, *How to Design AI for Social Good: Seven Essential Factors*, 2020 SCI. ENG. ETHICS 1771 (2020), https://doi.org/10.1007/s11948-020-00213-5.

Alternative corporate structures can enable the prioritization of user interests alongside financial returns. Public Benefit Corporations (PBCs) and similar forms provide a legal foundation for balancing multiple stakeholder interests, allowing BCI developers to explicitly prioritize neural privacy and cognitive autonomy in their corporate charters.[68] A BCI firm structured as a PBC would balance shareholder returns with commitments to mental privacy and user autonomy, reducing the risk that subtle manipulations or data-driven nudges override user welfare. These kinds of alternative corporate structures can attract investors specifically interested in ethical technology development while providing legal protection against shareholder pressures that might otherwise prioritize short-term returns over user wellbeing.[69]

Data stewardship models could also be leveraged to structurally separate neural data governance from profit-seeking functions. Data trusts and data cooperatives establish independent governance of user data with fiduciary obligations to data subjects rather than shareholders.[70] Under these models, neural data would be held by a separate entity bound by strict fiduciary obligations, while the commercial entity developing the BCI would access this data only under conditions aligned with user interests. The separation creates institutional barriers against the exploitation of neural data, even under financial pressure.

These corporate reforms can be encouraged through a combination of regulatory incentives, market differentiation, and investor pressure. Firms with robust AI governance attract both talent and investment, creating market incentives for fiduciary approaches even before regulatory requirements are fully implemented.[71] For brain foundation models, where user trust is essential for adoption, these market mechanisms may prove particularly powerful in driving corporate adoption of fiduciary standards.

Corporate governance reforms within individual jurisdictions, however, should be complemented by international coordination to prevent regulatory arbitrage and ensure consistent protection as brain foundation models increasingly operate across national boundaries.

### E. International Governance and Cross-Cultural Implementation

While fiduciary principles have deep roots in Western legal traditions, the global nature of neurotechnology development requires frameworks that can function across diverse legal, regulatory, and cultural contexts. Different regions approach the balance between individual rights, collective welfare, and technological governance in ways that reflect their distinct

---

philosophical, political, and cultural values. A robust fiduciary AI framework for brain foundation models would be adaptable while preserving core protections for mental autonomy.

The concept of fiduciary duties, loyalty, care, and confidentiality, finds analogues in multiple legal traditions, though their interpretation may vary. In East Asian jurisdictions influenced by Confucian values, for instance, fiduciary-like obligations often emphasize harmony and collective welfare alongside individual rights.[72] Japanese approaches to AI ethics, exemplified in their AI Governance Guidelines, focus on human-centered design while acknowledging broader social responsibilities.[73] These cultural frameworks could enrich Western fiduciary models by expanding the notion of best interest beyond individualistic conceptions to include familial and community considerations in how neural data is stewarded.

When deployed internationally, the tetraplegia assistance system described earlier would encounter varying cultural expectations about autonomy, requiring adaptable fiduciary frameworks that respect both universal principles and local norms. The practical implementation of fiduciary AI principles across diverse legal systems will require adaptive implementation frameworks, tiered compliance mechanisms, and cross-cultural dialogue.

Rather than imposing a single universal model, governance structures should identify core fiduciary principles for brain foundation models while allowing context-sensitive implementation. UNESCO's Ad Hoc Expert Group recommendations on neurotechnology (2024) and the OECD Framework for the Classification of AI Systems explicitly acknowledge cultural variation while establishing baseline protections.[74] Moreover, different jurisdictions have adopted varying levels of enforceability for AI governance, as evidenced by the contrast between the EU AI Act's binding regulatory approach and Japan's non-binding AI Social Principles. A tiered approach, from voluntary industry standards to binding regulations, can facilitate adoption across regions with different regulatory appetites while gradually harmonizing protections.

International forums connecting regulatory approaches across major neurotechnology markets have also begun to emerge. The OECD's Network of Experts on AI and the World Economic Forum's Global Futures Council on Neurotechnology represent structured efforts to create cross-cultural metrics for trustworthy neurotech governance.

---

[72] Nancy S. Jecker & Roger Yat-Nork, *Lessons from Li: A Confucian-inspired Approach to Global Bioethics*, J. MED. ETHICS 1 (2025), https://doi.org/10.1136/jme-2024-110480.

[73] Hiroki Habuka, *Japan's Approach to AI Regulation and Its Impact on the 2023 G7 Presidency*, CTR. STRATEGIC & INT'L STUD. (Feb. 2023), https://csis-website-prod.s3.amazonaws.com/s3fs-public/2023-02/230214_Habuka_Japan_AIRegulations.pdf?VersionId=BnLSQRRqoO9jQ8u1RW3SGKOA0i8DBc4Q.

[74] UNESCO, *Ethics of Neurotechnology: Towards an International Instrument*, https://www.unesco.org/en/ethics-neurotech/recommendation (accessed July 15, 2025); OECD, *OECD Framework for the Classification of AI Systems*, OECD DIGITAL ECONOMY PAPERS (Feb. 22, 2022), https://www.oecd.org/en/publications/oecd-framework-for-the-classification-of-ai-systems_cb6d9eca-en.html.

**Fiduciary AI for the Future of Brain-Technology Interactions**

While implementation paths may diverge, core concerns around human self-determination are shared.[75] Chile's constitutional protection of mental integrity, South Korea's inclusion of brain-machine interfaces in their Medical Devices Act, and emerging international efforts toward protecting mental privacy all reflect similar protective aims through different legal mechanisms.

As brain foundation models increasingly transcend national boundaries, with data potentially flowing across jurisdictions and users accessing services globally, these international considerations become more than theoretical concerns. The complex jurisdictional questions raised by neurotechnology developed and deployed across multiple legal frameworks require fiduciary models flexible enough to bridge diverse traditions while maintaining robust protections for neural data and cognitive liberty.

### F. Addressing Emerging Security Threats in Neural Technologies

The cross-border deployment of brain foundation models raises unprecedented national security and cognitive sovereignty concerns that traditional regulatory frameworks are ill-equipped to address. When BCI systems with foundation model capabilities operate across international boundaries, they create novel vulnerabilities that extend beyond conventional cybersecurity threats to the integrity of human cognition itself. These challenges require specialized governance approaches that complement the broader fiduciary framework while addressing unique security dimensions.

Critical security dimensions that require particular attention in an international fiduciary framework include the potential for cognitive hacking, data sovereignty across borders, and the potential militarization of BCI devices.

Cognitive hacking or cognitive warfare–which includes the deliberate manipulation of neural interfaces to influence thought processes or decision-making–represents an emerging frontier in information warfare. Research on neural influence techniques demonstrates that even subtle modulations of BCI feedback can systematically bias user decisions without conscious awareness.[76] A robust international fiduciary framework should establish clear boundaries between legitimate personalization and manipulative exploitation.

---

Data sovereignty questions will also become especially acute with neural data. When a BCI system trained in one jurisdiction processes the brain signals of users in another, complex questions arise about which fiduciary standards govern the interaction. The EU's approach of applying GDPR protections based on user location rather than company headquarters provides one model, but neural data requires additional safeguards given its potential to reveal cognitive states. These questions of neural data sovereignty will represent a novel challenge that requires governance frameworks that respect both national security interests and individual cognitive rights.

As military applications of BCI technology advance, from enhanced soldier performance to neural reconnaissance, the role of fiduciary principles in non-civilian contexts will become more crucial. International humanitarian law principles like distinction and proportionality could inform specialized fiduciary duties for military-grade BCIs, similar to how the Biological Weapons Convention established preventive norms before widespread deployment. The Uniform Law Commission's work on mental privacy and UNESCO's draft recommendations could serve as building blocks for an international convention specifically addressing cognitive security.

These cross-border challenges underscore why fiduciary AI frameworks must be embedded in broader international governance structures. Regional regulatory disparities could create disparate safeguards where entities deploy brain foundation models in jurisdictions with minimal fiduciary obligations, undermining protections globally. Coordinated international approaches, perhaps through specialized treaties or an international regulatory body focused on neurotechnology, will be essential to establish baseline fiduciary duties that transcend jurisdictional boundaries while addressing legitimate national security interests.

V.      **Challenges and Constraints in Implementing the Fiduciary AI Approach**

        A.   **Technical Challenges**

             a.   **Risks of Fiduciary AI Failures**

Despite its promise, an AI imbued with fiduciary duties is not infallible. One major risk is that the model might fail to uphold its obligations due to technical misalignment or unforeseen scenarios. AI systems, for example, can behave in unintended ways if their objectives are even slightly mis-specified – a classic alignment problem. It's hard to specify exactly what we want in code, and AI agents often find loopholes or shortcuts that deviate from our intent.[77] A model trained to maximize a certain health outcome might, unless carefully constrained, learn a strategy that technically achieves the metric but violates patient trust or wellbeing (the kind of gaming the system behavior seen in other AI contexts. In fiduciary terms, the AI could unintentionally prioritize a proxy goal over the user's true best interest, amounting to a breach of duty. There's also the risk of misgeneralization: a model might perform well in training (following fiduciary rules in typical cases) but then face a novel situation and err in a harmful way.

---

[77] Spencer Williams, *Layered Alignment*, 23 U. N.H. L. REV. (Mar. 6, 2025), https://dx.doi.org/10.2139/ssrn.5095543.

An AI mental health counselor, for example, might know not to disclose a patient's brain-scan-based diagnosis without consent, but if confronted with a law enforcement demand or a tricky social engineering attempt, it might not know how to respond and inadvertently leak protected information. Such failures could seriously erode user trust and cause harm the AI was supposed to prevent.

Mitigating these risks requires layers of safeguards. Robust training and testing are key: developers should subject fiduciary AI models to rigorous adversarial trials (simulating attempts to trick or misuse them) to ensure they consistently resist. This practice, known as red-teaming, is already used to probe AI for vulnerabilities.[78] Researchers have found that even well-intentioned models can sometimes be manipulated via cleverly crafted inputs to produce. For example, an experiment showed that different AI chatbots could be coaxed into giving advice for illicit activities by phrasing the prompt just right; some models refused, but others complied, highlighting the need for stronger internal guardrails.[79]

### b. Technical Barriers to Ethical Refusal

A key challenge in fiduciary AI is enabling systems to actively refuse unethical or non-fiduciary commands while remaining useful to the user. It is one thing for a chatbot to decline obviously disallowed content, but it is quite another for an AI that handles brain data to identify subtler ethical lines. The model should detect scenarios that violate its fiduciary duty, such as a request from an advertiser to access a user's EEG patterns for marketing. Large language models provide a starting point, some already refuse harmful instructions, but brain-focused AI faces more ambiguous situations. A further complication is that these models do not truly understand human values; they rely on training or constraints. Unfamiliar or re-framed unethical requests might slip past their refusal filters, as occurs with so-called jailbreaking techniques in today's chatbots. A malicious actor, for instance, could pose as the data subject and ask the AI to export raw neural data, exploiting any gaps in authentication or ethical reasoning.

At the same time, even a well-meaning AI can become excessively cautious. If it refuses every benign request to analyze or share brain signals, its protective reflex undermines its core purpose of assisting the user. Researchers have labeled this the false positive problem: restricting legitimate use out of fear of crossing an ethical boundary. Methods like multi-agent oversight, Constitutional AI, and fine-tuned RLHF (reinforcement learning from human feedback) may help the system maintain a middle ground, blocking truly unethical demands

---

[78] *Id.*
[79] *Id.*

while allowing beneficial data analysis in line with user intent.

These technical barriers to refusal intersect with broader ethical questions about user autonomy, often discussed under the umbrella of cognitive liberty. When AI systems are tasked with shielding users from harm, they risk becoming overprotective, effectively overruling an individual's choices. An AI entrusted with managing sensitive brain-data sharing, for example, might refuse a user's request if it deems the third-party recipient untrustworthy, even if the user has consciously decided the potential benefits outweigh the risks. Such paternalism can erode the user's right to shape their own mental domain and make informed decisions.

One response, drawn from Cass Sunstein's work, is the concept of soft paternalism. Rather than outright blocking the user's request, the AI warns about risks or suggests less invasive options. In doing so, it honors its duty of care without foreclosing user autonomy. Another complementary approach is dynamic consent. Here, the AI temporarily pauses actions if it detects confusion or pressure, offering clearer explanations or asking clarifying questions. If the user demonstrates genuine understanding, the AI proceeds, thus balancing the fiduciary obligation to protect the user with the user's right to decide how their brain data is used.

By weaving these features into fiduciary AI design, integrating technical refusal mechanisms with ethical safeguards against paternalism, developers can create systems that neither reflexively comply with every harmful demand nor unilaterally shut down user choices. The goal is to protect mental privacy and well-being without compromising individuals' fundamental liberty to direct their own brain data and cognitive experiences.

### B. Scalability Across Medical and Consumer Neurotechnology

In clinical and medical settings, fiduciary AI operates within well-defined ethical and legal frameworks, akin to those governing medical professionals. AI systems assisting in neurological diagnoses or running BCIs for medical use should comply with regulations such as the FDA's AI/ML framework in the United States or the European Union's Medical Device Regulation (EU MDR). In these contexts, fiduciary AI can integrate with established oversight mechanisms, ensuring ethical compliance while preserving patient autonomy.

By contrast, consumer neurotechnology, exemplified by wearable EEG headsets or cognitive-enhancement apps, often arises in unregulated or lightly regulated markets. Commercial incentives to monetize user data are especially strong here, as brain-based data can reveal highly sensitive insights that are tempting to resell or leverage for targeted advertising.

Balkin's concept of "information fiduciaries" offers a way to neutralize such exploitative uses: entities handling neural data would be legally bound to serve only the user's best interests, thwarting conflicts of interest. Under this model, a fiduciary AI agent could not simply collect or share brain signals for profit unless the user explicitly authorizes it for a clear, user-beneficial purpose.

Customizable fiduciary settings can further balance user autonomy with the AI's protective obligations. Individuals might allow their brain data to be shared for specific health research or personalized therapeutic tools yet prohibit its use for marketing or behavioral profiling. This tiered consent approach aligns with Farahany's (2023)[80] emphasis on adaptive governance, preserving core fiduciary duties while letting users fine-tune when and how their neural data is employed.

By tailoring fiduciary AI obligations to the specific hazards and purposes of each neurotechnology domain, governance models can remain both robust and adaptive. In doing so, they safeguard users' cognitive autonomy without stifling innovation, ensuring that fiduciary duties endure and scale effectively across diverse neurotech applications.

## VI. Conclusion

The advent of brain foundation models integrated into brain-computer interfaces (BCIs) marks a profound leap in artificial intelligence, moving beyond text or image analysis to real-time engagement with a user's mental states. At once, these new capabilities promise revolutionary benefits, from enhanced neuroprosthetics and advanced clinical diagnostics to brain-driven human–machine collaboration. But the same power that allows an AI to infer and adapt to unspoken thoughts can also subvert mental privacy, subtly steer cognition, or even rewrite neural pathways over time. It is precisely this capacity for intimate influence that demands a further-reaching response than conventional data privacy or AI risk-management frameworks typically provide.

A fiduciary AI paradigm, rooted in well-established legal doctrines such as the doctor–patient relationship, offers one way to meet these novel challenges. By formalizing duties of loyalty, care, and confidentiality in both software design and policy oversight mechanisms, we can require an agentic brain foundation model to act in alignment with its user's best interest. This is more than a set of abstract principles. It calls for a guardian model that checks for manipulative patterns in code to external audits that hold developers accountable for hidden incentives or corporate pressures. Critically, it also compels us to treat neural data not just as sensitive but as intimately reflective of individual self-determination, deserving of protections akin to those governing medical or legal confidences.

Such a fiduciary approach addresses the growing asymmetry in power between users and advanced brain–AI systems. As with other fiduciary relationships, doctor–patient, lawyer–client, the key insight is that certain interactions carry heightened vulnerability, requiring elevated safeguards. In BCIs, this means preventing covert exploitation of subconscious signals, blocking disloyal uses of brain data, and allowing for dynamic consent or user override features that preserve cognitive liberty. By adding legal mandates to the technical guardrails explored in the paper, like explainability, sandboxing, and multi-tiered oversight, societies can anchor the constitutional AI concept in enforceable norms that make both corporate and governmental actors accountable for any breach of mental autonomy.

---

[80] NITA A. FARAHANY, THE BATTLE FOR YOUR BRAIN: DEFENDING THE RIGHT TO THINK FREELY IN THE AGE OF NEUROTECHNOLOGY (2023).

Far from stifling innovation, these fiduciary obligations can foster public trust and drive ethically aligned business models. In fields as diverse as precision medicine, mental health support, and next-generation user interfaces, assurances that AI systems can uphold user welfare can expand legitimate research and adoption without succumbing to the pressures of "surveillance capitalism." Ongoing developments in rights over neural data and integrity, data privacy expansions, and global AI rulemaking (e.g., the EU AI Act) create fertile ground for integrating fiduciary clauses aimed specifically at BCI with integrated brain foundation models.

BCIs that can make decisions for users also raise important questions across multiple fields. These intimate technologies inevitably reflect their creators' values and assumptions. This raises a question about whether ethical guidelines designed for chatbots work for devices wired directly into our brains? Generic approaches may miss the specific needs of actual users. We also need to decide how much control users should have over their devices—should people be able to customize their brain interface like adjusting settings on a phone? While user control respects personal choice, it conflicts with the reality that brain implants have limited computing power and strict safety requirements. Addressing these challenges will require collaboration between neuroscientists, engineers, ethicists, and the people who use these devices.

Ultimately, the core proposition is simple yet groundbreaking. If an AI system can interpret or reshape the human mind in real time, it should be legally and ethically bound to advance the user's self-determination above all else. Only in this manner can the extraordinary promise of agentic BCIs, to restore, augment, and deeply enrich human capacities, unfold without sacrificing the foundational right to one's own cognitive liberty. By placing brain foundation models on fiduciary footing, we take an essential step toward ensuring that the next leap in AI-driven neurotechnology remains a force for human empowerment.